# On the State of Coherence in the Land of Type Classes

Dimi Racordon[a] , Eugene Flesselle[a] , and Cao Nguyen Pham[a]

a    LAMP, EPFL, Switzerland

**Abstract**    Type classes are a popular tool for implementing generic algorithms and data structures without loss of efficiency, bridging the gap between parametric and ad-hoc polymorphism. Since their initial development in Haskell, they now feature prominently in numerous other industry-ready programming languages, notably including Swift, Rust, and Scala. The success of type classes hinges in large part on the compilers' ability to infer arguments to implicit parameters by means of a type-directed resolution. This technique, sometimes dubbed *implicit programming*, lets users elide information that the language implementation can deduce from the context, such as the implementation of a particular type class.

One drawback of implicit programming is that a type-directed resolution may yield ambiguous results, thereby threatening coherence, the property that valid programs have exactly one meaning. This issue has divided the community on the right approach to address it. One side advocates for flexibility where implicit resolution is context-sensitive and often relies on dependent typing features to uphold soundness. The other holds that context should not stand in the way of equational reasoning and typically imposes that type class instances be unique across the entire program to fend off ambiguities.

Although there exists a large body of work on type classes and implicit programming, most of the scholarly literature focuses on a few select languages and offers little insight into other mainstream projects. Meanwhile, the latter have evolved similar features and/or restrictions under different names, making it difficult for language users and designers to get a sense of the full design space. To alleviate this issue, we set to examine Swift, Rust, and Scala, three popular languages featuring type classes heavily, and relate their approach to coherence to Haskell's. It turns out that, beyond superficial syntactic differences, Swift, Rust, and Haskell are actually strikingly similar in that the three languages offer comparable strategies to work around the limitations of the uniqueness of type class instances.

## The Art, Science, and Engineering of Programming

## 1 Introduction

Type classes [31] have become a popular tool to practice generic programming, the discipline of lifting general abstractions from concrete implementations without loss of efficiency [15]. A type class is a set of requirements representing a *concept*, in the form of operations that a type must support. The conformance of a type *T* to a concept *P* is then *witnessed* by an object implementing all of the *P*'s requirements—i.e., a *model* of *T*'s conformance to *P*—that is used in generic code.

Listing 1 shows an example in Haskell. We define a concept describing the well-known iterator pattern [9]. This concept requires a single operation, `next`, which returns either an element and an advanced iterator, or the absence of any value if the iterator reached the end of the sequence. This operation involves an associated type [3, 4], abstractly representing the type of elements produced by the iterator. We then define a generic reduction operator with a constraint on its type parameter, written on the left of `=>`, stating that the function can only be applied to types conforming to `Iterator`. The constraint further lets us write `Element a` to refer to the associated type of a. Next, we define a model witnessing how `Word64` implements `Iterator`, using 64-bit unsigned integers as null-terminated buffers of unsigned bytes. Finally, we apply our generic `fold` to sum the contents of a buffer storing 2 bytes.

■ **Listing 1**   Generic programming with concepts

```
1  class Iterator self where
2    type Element self
3    next :: self -> Maybe (Element self, self)
4
5  fold :: Iterator a => a -> b -> (b -> Element a -> b) -> b
6  fold xs x f = case (next xs) of
7    Just (y, ys) -> fold ys (f x y) f
8    Nothing -> x
9
10 instance Iterator Word64 where
11   type Element Word64 = Word8
12   next 0 = Nothing
13   next n = Just (fromIntegral (n .&. 0xff), n shiftR 8)
14
15 main = do print (show (fold (0x2a2a::Word64) (0x00::Word8) (+)))  -- Prints "84"
```

One benefit of concept-based programming over more classical approaches such as subtyping is the ability to *retroactively*—i.e., non-invasively [32]—define the conformance of a type to a particular concept. In Listing 1, for example, we stated that unsigned integers can behave like iterators without modifying the definition of `Word64`. In contrast, if `Iterator` was defined as a class to be inherited in an object-oriented language, we would not be able to establish `Word64` as a subclass of `Iterator` without modifying the standard library. Instead, we would be compelled to define an adapter, resulting in cumbersome boilerplate to forward operations to the wrapped instance. We illustrate this point further in Section 2.

Retroactive conformance requires that an ad-hoc model relating concept requirements to their implementations be passed to generic algorithms. Although this model could be passed explicitly—`fold` could accept a witness of a's conformance to `Iterator` as an explicit parameter—the ergonomics of concepts rest on the compiler's ability to hide this process. In other words, the compiler must infer arguments to implicit parameters [16]. That is straightforward in Listing 1: as a is instantiated to `Word64`, the compiler is prompted to find a model of `Word64`'s conformance to `Iterator` to satisfy the function's constraint, which is immediately found. As we will see however, implicit resolution is in general a complex process that even becomes undecidable when combined with typically desired features of concept systems, such as non-syntactic type equalities [29] and recursive conformance constraints [17]. In fact, looking for a model boils down to a type inhabitation problem, which is known to be undecidable with universal quantification [7, 25].

Another compounding problem is that implicit resolution may be ambiguous in the presence of overlapping model definitions. Hence, maintaining *coherence*—the property that valid programs have exactly one meaning [20]—involves additional mechanisms to reject ambiguities. The proper way to address this issue has been at the center of an ongoing debate. On one side, proponents of dependent typing advocate for flexible systems where implicit resolution is context sensitive, with models being first-class objects governed by the same rules as other values. On the other side, equational reasoning is put front and center, thus restricting the influence of context-dependent information on implicit resolution; models are usually second-class values with extra rules to handle associated types and non-syntactic equalities.

The second approach traditionally requires that a program defines at most one unique model of a given type to a given concept. This limitation trivially guarantees that implicit resolution cannot produce ambiguous results or non-deterministically select a model when another could have been found by expanding the search space. Unfortunately, this strategy, also known as *global uniqueness of type class instances* [35], hinders expressiveness. For instance, it implies that integers may not be seen as monoids under multiplication in one part of the program and under addition in another. Even more concerning, it works against modular software design, since defining an instance in a library may prevent interoperability with any piece of software with potentially overlapping instances, regardless of their same or differing definitions. While formal calculi may get away with such a restriction, real-world programming languages typically attempt to relax it and/or provide workarounds. Alas, doing so results in convoluted systems—Haskell's twelve extensions related to type classes or Rust's orphan rules [21]—and may open soundness holes [33].

The state of the art surrounding this issue is rich but hard to navigate. Although there exists a vast body of work on type classes in Haskell, far less ink has been spilled studying other mainstream languages, which have evolved similar features under different names and developed competing approaches under different constraints. We posit that this situation makes it difficult for language designers to get a clear view of the complete design space, increasing the chance to rediscover past results or fall into previously uncovered traps, as evidenced by an unsoundness issue in Swift that this work revealed—we discuss it in Section 3.2.2.

To alleviate this issue, we set to explore the state-of-the-art to describe the current standing of techniques aimed at upholding coherence. Specifically, we look at Swift,

Rust, and Haskell to examine how each of them works within or around the limitations of global model uniqueness and identify two contentious design questions:

1. Where and how should overlapping model definitions be detected to guarantee global model uniqueness?

2. How can library authors define models without causing downstream source breakage due to overlapping conflicts?

We then study these questions in the context of Scala, a language offering a more flexible approach to concept-based programming that leverages path dependent types [1]. Our work does not attempt to elect a "best language". Our investigation focuses on global model uniqueness, questioning whether it is actually the right path to uphold coherence, even in non-dependently typed languages.

## 2  Background

Type classes were first introduced in a paper titled "How to Make *ad-hoc* Polymorphism Less *ad-hoc*." [31] This title alludes to a reconciliation of the two forms of polymorphisms introduced by Strachey [28]: parametric polymorphism, in which generic operations behave uniformly, notwithstanding the type of their inputs, and *ad-hoc* polymorphism, in which the compiler uses some mechanism—e.g., overload resolution—in addition to its type system to choose between different unrelated operations. Both approaches have their merits. In particular, *ad-hoc* polymorphism better supports behavioral extensibility and can sometimes outperform its counterpart by reducing abstraction penalty [6]. On the other hand, parametric polymorphism composes better and is more amenable to definition site checking [18]. Type classes bridge the gap between these approaches by bundling ad-hoc operations in an object that is passed to parametric abstractions. To illustrate the motivation, let us revisit our example from Listing 1, but with explicit witness passing:

■ **Listing 2**  Generic Programming without Concepts

```
1   fold_alt :: (a -> Maybe (b, a)) -> a -> c -> (c -> b -> c) -> c
2   fold_alt next xs x f = case (next xs) of
3     Just (y, ys) -> fold next ys (f x y) f
4     Nothing -> x
5
6   nextByte :: Word64 -> Maybe (Word8, Word64)
7   nextByte 0 = Nothing
8   nextByte n = Just (fromIntegral (n .&. 0xff), n `shiftR` 8)
9
10  main = do print (fold_alt nextByte 0x2a2a 0 (+)) -- Prints "84"
```

Again, fold_alt is a generic reduction operator. This time around, however, no constraint is placed on its generic parameters. Instead, the operation has one more generic type parameter and accepts an additional argument representing the next operation of the Iterator class we defined earlier. This operation is implemented by nextByte in line 6, making 64-bit unsigned integers notionally "conform" to the concept that the Iterator class represents.

Listing 2 demonstrates that System F is sufficient to (weakly [8]) express generic algorithms using concepts. While `fold_alt` remains acceptable, as `Iterator` defines a single operation and associated type, Siek and Lumsdaine [24] observe however, this approach does not scale in practice because more complex algorithms may require dozens of type-specific operations to be documented and passed down pervasively. Concepts eliminate the tedious argument plumbing, while also simplifying the signatures of generic algorithms. In the above program, the relationship between `a` and `b` is implicit whereas the one between `a` and its associated type `Element a` in Listing 1 is explicitly captured in a reusable abstraction that the user may consult.

The remainder of this section introduces some important definitions related to type classes. Although we use Haskell to illustrate examples, the notions we discuss apply broadly to programming languages supporting concept-oriented programming.

### 2.1 Constraints

Type classes are a way to place *constraints* on generic parameters. For example, in Listing 1, `a` is constrained to implement `Iterator`'s requirements. Such a constraint occupies two roles [18]. At definition site, it provides assumptions to type check the body of `fold`. At use site, it serves as preconditions that are mechanically verified. That is unlike type validation, such as C++20's concepts [30], where definition sites are optimistically assumed correct until type variables are instantiated.[1]

Two kinds of constraints can be used to specify generic operations: *conformance constraints* require a model of a type's conformance to a particular concept and *equality constraints* require that two types be equal. The following program demonstrates these two kinds in action:

■ **Listing 3**   Conformance and equality constraints

```
1  elementsEqual :: (Iterator a, Iterator b, Eq (Element a), Element a ~ Element b) => a -> b -> Bool
2  elementsEqual xs ys = case (next xs, next ys) of
3    (Just (l, ls), Just (r, rs)) | l == r -> elementsEqual ls rs
4    (Nothing, Nothing) -> True
5    otherwise -> False
```

Here, `elementsEqual` accepts two iterators and returns whether they produce the same elements in the same order. Four constraints on the context in which the abstraction is applicable are specified. The first two are conformance constraints ensuring that `a` and `b` are indeed iterators. The third states that the elements produced by the first iterator can be compared for equality. Finally, the equality constraint `Element a ~ Element b` states that the two iterators produce elements of the same type.

Equality constraints invite a collection of complications into type checking, which Schrijvers et al. dub the *entailment problem* [22]: given a signature, prove that some type equalities are implied by its constraints. The problem is further compound by type inference, which allows compilers to deduce type arguments from context. For instance, given the call elementsEqual (0x2a::Word64) [] and assuming lists conform to Iterator, a competent compiler will deduce that the second argument is a list of Word8.

---

[1] Concepts found in modern C++ do not match the definition we use in this paper as they only constrain use sites. The name was inherited from an earlier proposal [11] that was in line with our terminology, which we borrow from Stepanov's [27].

## 2.2 Polymorphic Models

The synergy between type classes and parametric polymorphism extends to model definitions, which may also be polymorphic. Specifically, we can exploit parametric polymorphism to describe how a variety of different types conform to a concept in a regular way. For example, Listing 4 states that all applications of the type constructor Maybe conform to Iterator. Here, no constraint is placed on the model's definition since conformance can be established for any a. Listing 5 does similarly for half-closed ranges. Here, the conformance can only be established if there is a way to "increment" the lower bound, which is expressed by placing a constraint on the model's generic parameter. In both cases, the declaration of these polymorphic models let us re-use generic algorithms like fold from Listing 1.

■ **Listing 4**   Polymorphic conformance

```
1  instance Iterator (Maybe a) where
2      type Element (Maybe a) = a
3      next Nothing = Nothing
4      next (Just x) = Just (x, Nothing)
5
6  main = do print (show (fold (Just 42) 0 (+)))  -- Prints "42"
```

■ **Listing 5**   Conditional Conformance

```
1  data Range a = a `UpTo` a
2  instance Integral a => Iterator (Range a) where
3      type Element (Range a) = a
4      next (x `UpTo` y)
5          | x < y = Just (x, (x + 1) `UpTo` y)
6          | otherwise = Nothing
7
8  main = do print (show (fold (1 `UpTo` 4) 0 (+)))  -- Prints "6"
```

## 2.3 Polymorphic Type Classes

All type classes presented so far are parameterized by a single variable, representing their conforming types. We say that a type class is polymorphic when additional variables are used to abstract over other aspects of a concept. Models of polymorphic type classes can define conformance to either a specific application or remain generic, so as to define how a type conforms to a family of related concepts.

In the program below, Convertible describes values that can be converted to instances of another type. The two instances illustrates different ways to conform: the first states that an Int can be converted to a String whereas the second states that any value can be wrapped in an optional.

■ **Listing 6**  Polymorphic type class

```
1  class Convertible a b where
2    convert :: a -> b
3
4  instance Convertible Int String where
5    convert n = show n
6  instance Convertible a (Maybe a) where
7    convert a = Just a
```

Polymorphic type classes let us lift even more general abstractions from multiple implementations. For instance, we can further generalize `elementsEqual` so that it can apply to any pair of iterators, provided that the elements produced by the former can be converted to the type of the elements produced by the latter.

■ **Listing 7**  Application of a polymorphic type class

```
1  elementsEqualWithConversion ::
2    (Iterator a, Iterator b, Eq (Element a), Convertible (Element a) (Element b)) =>
3    a -> b -> Bool
4  elementsEqualWithConversion xs ys = case (next xs, next ys) of
5    (Just (l, ls), Just (r, rs)) | l == convert r -> elementsEqualWithConversion ls rs
6    (Nothing, Nothing) -> True
7    otherwise -> False
```

Polymorphic type classes are not available in Swift and are not standard Haskell (requires the `MultiParamTypeClasses` extension and others to be fully usable). However, they are used extensively in Rust, notably in the standard library with the `From<T>` concept—which is similar to `Convertible`—and binary operator overloading concepts. In Scala, polymorphic type classes are naturally supported by being first-class objects.

## 2.4  Concept Refinement and Specialization

A concept $C_2$ *refines* another concept $C_1$ if whenever a type conforms to $C_2$ it also conforms to $C_1$ [27]. For instance, we could say that if a type has a total order, then it is also comparable for equality. If each of these two notions is expressed with a corresponding concept, then the former will refine the latter. Languages featuring type classes can typically capture this idea in the form of a class hierarchy where subclasses denote refinements of their superclasses.

■ **Listing 8**  Concept refinement

```
1  class Equatable self where
2    (==) :: self -> self -> Bool
3  class (Equatable self) => Comparable self where
4    (<) :: self -> self -> Bool
```

Although the terminology is reminiscent of subtyping, another form of polymorphism, concept refinement does not necessarily convey the same notion. In particular, refinement does not typically entail substitutability as proposed by Liskov and Wing [13]: just because `Equatable` refines `Comparable` does not mean a model of the latter can be used in places where a model of the former is expected in general.

Nonetheless, refinement hints the ability to *specialize* an algorithm depending on the characteristics of its operands. For instance, imagine we define an operation to determine whether an iterator produces a palindrome. Such a test can be implemented using our Iterator concept and a local stack. However, a refinement describing iterators able to yield elements from both ends would let us dispense with the stack and write a more efficient algorithm. It could be desirable to have the language implementation "dispatch" to the more efficient implementation whenever the iterated sequence is double-ended. Although Haskell does not support such a mechanism, specialization is common in other programming languages, such as Swift and C++.

## 2.5  Implicit Programming

Using type classes to describe abstract concepts is only half the story. Another important component is the ability to "eliminate the tedious plumbing" involved in passing type-specific operations around. In Listing 2, that means having the compiler infer arguments for the parameters b and nextByte. Nearly all programming languages featuring parametric polymorphism can also infer arguments to type abstractions. Implicit programming consists of doing the same for *term* abstractions.

Implicit programming is a type-directed process. When an argument must be inferred, the compiler either looks for a value having the right type in the current implicit scope or attempts to construct one using type-directed rules. For instance, in Listing 5, the call fold (1 `UpTo` 5) 0 (+) requires the compiler to resolve a model witnessing the conformance of Range Int to Iterator. No such model exists in the implicit scope but the instance definition in line 2 will produce one, given a conformance of Int to Integral, prompting the compiler to look for it. Such a conformance can be fetched from the standard library, and so the process concludes.

The semantics of a program relying on implicit resolution are usually given in terms of *elaboration*. Type checking a term of the source language results in a translation to a simpler language—e.g., System F—where models are dictionaries or records mapping a concept's requirements to their respective implementations. Examples of elaboration semantics include Scala's implicits [16], Agda's instance arguments [5], OCaml's modular implicits [34], and concepts in Siek's System F$^G$ [24].

Dictionary passing leaves dependently typed languages at an advantage when associated types are considered because their type systems are already powerful enough to express type selections on terms. For instance, the associated type of Word64's conformance to Iterator in Listing 1 can be expressed as a type member in DOT [1]. In contrast, non-dependently-typed languages require more sophisticated elaboration schemes to translate associated types without threatening soundness.

## 2.6  Coherence

As mentioned in the introduction, one thorny problem posed by implicit programming is to guarantee that valid programs have only one meaning—i.e., their semantics are not ambiguous—a property also known as *coherence* [20]. That is, to rephrase once more, any well-typed term must have unambiguous evaluation semantics. For instance, suppose we compose the program in Listing 1 with a piece of code in which we defined another conformance of `Word64` to `Iterator`. Any call to `fold` would then become ambiguous. Implicit resolution could find two equally valid models and therefore be unable to determine how a particular call site intends to use `Word64`.

Possible sources of incoherence are not always as obvious. To illustrate, consider the program below, which extends Listing 1. We define a concept `StringConvertible` describing values having a textual representation and three models. The first two concretely apply to `Word64` and `Word8`, respectively. The third applies to any iterator whose elements also conform to `StringConvertible`, by simply iterating through the sequence to produce the textual representation of each element.

■ **Listing 9** Overlapping Instances

```
1  class StringConvertible a where
2    toString :: a -> String
3
4  instance StringConvertible Word64 where
5    toString n = show n
6
7  instance StringConvertible Word8 where
8    toString n = show n
9
10 instance (Iterator a, StringConvertible (Element a)) => StringConvertible a where
11   toString xs = "[" ++ (fold xs "" (\s -> \x -> s ++ (toString x) ++ ",")) ++ "]"
12
13 main = do print (toString (0x2a2a::Word64))  -- Prints either "[42,42,]" or "10794"
```

The model definition in line 10 of Listing 9 may seem like a reasonable way to eliminate some of the boilerplate that would otherwise be involved in defining conformances to `StringConvertible` for every iterator.[2] Unfortunately, it can also cause incoherence because there are now two ways to resolve `StringConvertible` in line 13. One applies the definition in line 4 and the other applies the definition in line 10, satisfying the `Iterator Word64` requirement with the conformance from Listing 1.

As we discuss more extensively in Section 3.1, there are essentially two ways to reject programs like the one Listing 9 in order to uphold coherence of model resolution. The first is to *guarantee* that a pair of instances may never overlap, by checking this property at each model's definition site. In the above example, that would mean rejecting the model definition line 10 because there exists an instantiation of a that would overlap. The second is to *verify* at each use site that only one model can be resolved. In the above example, that would mean rejecting the call to `toString` because the compiler can resolve two possible models. An extension of this approach may apply some prioritization scheme to disambiguate competing candidates. For example, it would be reasonable to select the conformance of `Word64` to `Iterator` that is defined in the standard library because it matches more "easily"—i.e., without additional deduction steps—the requested model.

---

[2] Another strategy would be to apply Haskell's deriving mechanism [2] on each iterator type. However, this approach would not support retroactive modeling.

Similarly, polymorphic type classes can introduce ambiguities: a call to convert (3 :: Int) can be resolved to either instances defined in Listing 6. However, these are often treated as a type inference problem rather than one of coherence.

### 2.7 Stability

A popular and effective way to understand operational semantics is to think in terms of substitutions, a practice also known as *equational reasoning*. For instance, to reduce a type abstraction in System F, we can simply remove the binder and substitute the type argument for the type variable in the abstraction's body:

$$(\Lambda\alpha.\lambda x : \alpha \to \alpha.\lambda y : \alpha.x\,y)\,\mathbb{B}\ not\ true \longrightarrow (\lambda x : \mathbb{B} \to \mathbb{B}.\lambda y : \mathbb{B}.x\,y)\,not\ true$$

It is often desired that substitution preserve semantics—i.e., substitution of equals does not change the semantics—especially in the context of functional programs. Such a property is upheld in the above example: the term on the left side of the arrow has the exact same meaning as the one of the right, in the sense that they are contextually equivalent [14]. Languages satisfying this property are said to be *stable under type substitution* [23].

Stability can be lost in the presence of overlapping models because instantiating type variables may modify the result of implicit resolution. To illustrate, imagine that we add the following function to Listing 9, which defined models of StringConvertible for iterators producing elements having a textual representation:

■ **Listing 10**   Unstable Definition

```
1  log :: a -> IO ()
2  log m = print (toString m)
```

The call to toString causes the compiler to look for a conformance to StringConvertible. Given the available information, the compiler's only solution is to use the generic model in Listing 9. However, instantiating log's type parameter at Word64 would present an additional candidate, thus leading to unstable resolution.

Note that, though stability relates to coherence, the two notions deserve a distinction. In particular, the latter can hold without the former. In fact, most programming languages with both subtyping and overloading are typically coherent but unstable since narrowing may refine overload resolution.

### 2.8 Approaches to Coherence

Implicit programming shelters two competing doctrines. The first holds that context should not stand in the way of equational reasoning, which helps understanding, refactoring, and optimizing programs. The second holds that context ought to be local and flexible lest modular design becomes impractical.

Incoherence stems from having ambiguous choices. Hence, a radical way to get coherence is to ensure the uniqueness of any model witnessing the conformance of particular type to particular concept, a strategy also known as *global uniqueness of type class instances* [35]. Stability follows since no amount of inlining can modify the result of implicit resolution: if a model is found, then there was only one to be found. Unfortunately, though this strategy looks very appealing on paper, it causes friction in practice as we shall see in the following section.

Conversely, proponents of dependent typing, such as Scala [16] and Coq [26], typically advocate for models being first-class objects governed by the same rules as other values, i.e. without scoping or overlapping restrictions. Implicit resolution is then merely a way to omit certain arguments and immediately benefits the full flexibility and soundness properties from using a powerful dependently-typed language. While these systems are far more permissive in the allowed derivations, they typically involve intricate disambiguation policies and lack completeness guarantees for the inference of these derivations.

## 3   Coherence by Global Model Uniqueness in Practice

The minimum necessary condition to uphold coherence is that implicit resolution be unambiguous. As mentioned, global model uniqueness is sufficient to guarantee this property. Nonetheless, two important questions that arise in programming languages adopting this strategy:

1. Where and how should overlapping model definitions be detected to guarantee global model uniqueness?

2. How can library authors define models without causing downstream source breakage due to overlapping conflicts?

We discuss these questions in the context of Haskell, Swift, and Rust, three languages that rely on global model uniqueness to uphold coherence and have overall similar approaches to type classes beyond superficial syntactic differences. The two programs below illustrates our running example expressed in Swift and Rust. Note that in both languages the type parameter representing the conforming type is called `Self` and is elided from protocol and trait declarations.

*Swift*

```
1  protocol Iterator {
2      associatedtype Element
3      func next() -> (Element, Self)?
4  }
5
6  extension UInt64: Iterator {
7      typealias Element = UInt8
8      func next() -> (UInt8, UInt64)? { ... }
9  }
```

*Rust*

```
trait Iterator: Sized {
    type Element;
    fn next(self) -> Option<(Self::Element, Self)>;
}

impl Iterator for u64 {
    Element = u8;
    fn next(self) -> Option<(u8, u64)> { ... }
}
```

### 3.1   Detecting Overlapping Models

To introduce the problem, consider the following example, which presents an overlapping model definition similar to the one discussed in Listing 9:

■ **Listing 11** Unambiguous and ambiguous resolution

```
1  -- From standard library: Int and Double conform to Show
2
3  instance Show a => StringConvertible (Maybe a) where
4    toString (Just x) = show x
5    toString Nothing = "nothing"
6
7  instance StringConvertible (Maybe Int) where
8    toString (Just x) = show x
9    toString Nothing = "NaN"
10
11 main = do
12   print (toString (Just (1.0 :: Double))) -- Unambiguous
13   print (toString (Just (1 :: Int))) -- Ambiguous
```

The first model definition states that `Maybe a` conforms to `StringConvertible` for any type a that conforms to `Show`, which include `Int` and `Double` from the standard library. The second model definition states that `Maybe Int` conforms to `StringConvertible`, unconditionally. The resolution at line 12 of `Maybe Double`'s conformance to `StringConvertible` is unambiguous. At line 13 however, there is not a unique model of `Maybe Int`'s conformance, since both definitions apply. We say that they *overlap*: there exists an instantiation of their free type variables such that they provide a model of the same type, without considering their requirements. The remainder of this section discusses the different ways languages adopting global model uniqueness attempt to detect these kind of ambiguities.

There are two places where overlaps can be detected, namely at definition site or at use site. As a program typically contains more uses than definitions, the first option has the advantage to reduce the possible errors that can stem from a problematic declaration. On the other hand, this approach is overly restrictive. In Listing 11, for example, the second model definition would be rejected, thus preventing not only the ambiguous resolution at line 13, but also the unproblematic instantiation at line 12.

Haskell adopts the second option, verifying at each use site that a single model is applicable. As a result, it accepts programs with overlapping definitions similar to Listing 11 with one caveat: the compiler must be able to tell definitions apart without considering constraints on generic parameters. For instance, the following definition is rejected, being considered a duplicate of the one in line 3:

```
1  instance StringConvertible a => StringConvertible (Maybe a) where -- <- rejected as duplicate
2    toString (Just x) = toString x
3    toString Nothing = "nothing"
```

Conflicting—but non-duplicate—definitions can coexist as long as no expression is causing implicit resolution to detect any ambiguity. In this case, the only recourse is to compile the program with the `IncoherentInstances` extension [10]. Unfortunately, that is not possible if the conflicting definitions lie in separate compilation targets. We discuss this situation in more details in Section 3.2.

Swift and Rust adopt the first option, checking for overlaps at definition sites. This strategy is harder to implement since model uniqueness mandates a check that generic models can never be instantiated in such a way that conflicts occur. Put differently, one must prove that the sets of models begotten by the two definitions are disjoint, which boils down to type inhabitation. Some restrictions can be introduced to keep the problem decidable but no disambiguation scheme can preserve stability [12].

Swift does not allow introducing arbitrary type variables in model definitions, requiring that they appear as argument of some type abstraction. For instance, while one can state that `Optional<A>`—which is analogous to `Maybe a` in Haskell—conforms to some concept Q if `A` conforms to `P`, one *cannot* state that any type conforming to `P` also conforms to Q. We presented a use case for such a conformance in Listing 9. Second, Swift does not admit more than one model definition for any given pair of a concept and a type constructor, which works as a—very strict—disjointness check. The following snippet illustrates:

```
1  extension Optional: StringConvertible where Wrapped: Show {
2    var toString: String { self.map(\.show) ?? "nothing" }
3  }
4  extension Optional<Int>: StringConvertible { // <- conflicting model
5    var toString: String { self.map(\.description) ?? "NaN" }
6  }
```

The first definition is legal and describes the conformance of `Optional<Wrapped>` to `StringConvertible`. Note that the application of `Optional` to `Wrapped` is implicit, based on the fact that the generic parameter of the type constructor is named `Wrapped`. The second definition is rejected because it may overlap with the first one. Perhaps surprisingly, the conflict is reported notwithstanding the lack of a conformance of `Int` to `StringConvertible`. That is due to the way Swift desugars type applications. Specifically, the second definition is interpreted as follows:

```
1  extension Optional: StringConvertible where Wrapped == Int {
2    var toString: String { self.description ?? "nothing" }
3  }
```

Given this desugaring, the compiler sees two definitions having requirements which it will not examine further. This limitation is rather severe as it implies that one cannot provide conformances to a concept for two distinct applications of a single type abstraction, even if those applications do not actually require quantification. For example, one can state that `Int` and `Bool` are both `StringConvertible` but one *cannot* do the same for `Optional<Int>` and `Optional<Bool>` via two declarations.

Unlike Swift, Rust will accept the above program if and only if it can prove that `Int` does not conform to `Show`. Such a proof is tractable if either of the bounds maps generic parameters to concrete types. In this case, the compiler can compute the sets of concepts to which these concrete types conform and check whether they intersect with the concepts required by the other model. In the absence of concrete types, however, Rust conservatively assumes there exists an inhabitant satisfying both bounds:

```
1  impl<T: Show> StringConvertible for Option<T> {
2    fn to_string(&self) -> String { self.as_ref().map_or("nothing".to_owned(), |v| v.show()) }
3  }
4  impl<T: Display> StringConvertible for Option<T> { // <- conflicting model
5    fn to_string(&self) -> String { self.as_ref().map_or("nothing".to_owned(), |v| format!("{}", v)) }
6  }
```

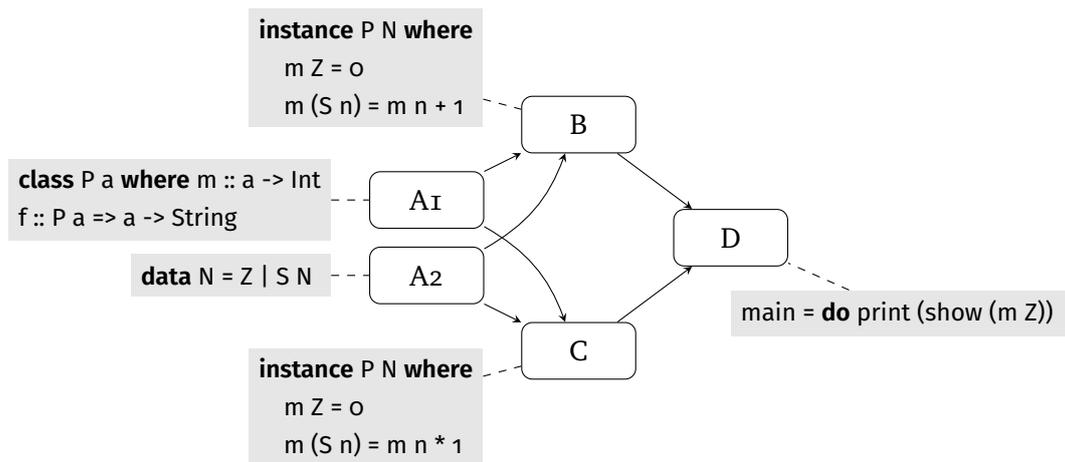

■ **Figure 1** Dependencies graph of a modular program.

We note that this restriction could be further alleviated to accept even more unproblematic situations. Since the problem is ultimately undecidable, though, the design tradeoff is between the complexity brought by more elaborate rules and the over-conservativeness of the implementation.

Since Haskell does not attempt to prove disjointness for arbitrary bounds, we observe that delaying the detection of overlapping models until the point of use does buy significantly more flexibility. In effect, Haskell imposes the same restrictions as Rust at the point of definition. As a result, its lazy approach to overlap detection is questionable, as it suffers similar problems as use site checking with respect to user experience [18]. In particular, a user may stumble on an incoherent resolution even if they did not authored the conflicting definitions. Hence, early detection may be preferable unless arbitrary definitions can coexist and/or the user can apply some explicit disambiguation mechanism to fix conflicts at use site.

### 3.2 Coherence Across Libraries

A crippling issue of model uniqueness relates to modular design. To illustrate, consider the diagram in Figure 1, which depicts the dependency graph of some hypothetical software. Rectangles denote encapsulated software components meant to be developed separately, the contents of these components is shown in gray boxes, and an arrow from $x$ to $y$ means that $y$ uses definitions from $x$. The diagram can be understood at different scales. Components could be individual libraries (e.g., a *module* in Haskell and Swift or a *crate* in Rust) or different pockets of code co-existing in the same file.

As we have mentioned in the introduction, one benefit of type classes is the ability to retroactively define new conformances. This benefit is at play here, allowing $B$ to reuse the generic algorithm f, defined in $A1$, with an instance of $N$, defined in $A2$. Unfortunately, doing so leaves $B$ incompatible with any other component pulling the same trick. That is the case here, as $C$ has also defined a conformance of $N$ to $P$. As a result, $D$ cannot depend on both $B$ and $C$, otherwise the program as a whole would violate model uniqueness.

This situation can be particularly vexing if the author of *D* is not also the author of *B* and *C*. There is nothing the author of *D* can do to fix the ambiguity. Furthermore, there is nothing the author of *B* could have done to predict the ambiguity. As the compiler processes *B*, it has no way of knowing the model definition in this module will eventually overlap with the one from *C*. More generally, unless the compiler can see the entire program at once, it cannot possibly tell whether a new definition may cause overlapping elsewhere downstream. The fundamental problem revealed by this example is that global uniqueness works against modular design as adding new models might break someone else's code.

In fact, Figure 1 depicts two *orphan instances*. A model witnessing the conformance of *T* to *P* is an orphan when it is defined in a module that contains neither defines *T* nor *P*. Such a model should generally be avoided, specifically because it can cause incoherence downstream. This problem is aggravating because there is nothing the downstream module—*D* in the above example—can do to fix the situation, as models are anonymous. Hence, there is no way to exclude an orphan from the implicit scope once its containing module has been imported.

Among the three languages we have examined, Rust is the only one addressing this issue, by means of its infamous *orphan rules* [21]. Orphan instances can cause compiler crashes in Swift and lead to incoherence programs in Haskell [35].

### 3.2.1 Rust's Orphan Rules

In a nutshell, the orphan rules prescribe that a model can be defined only in the crate that also defines either the type or the concept to which it relates. Thanks to this restriction, it is sufficient to only consider the definitions that the compiler can see while checking for possible overlaps, applying the strategy discussed in Section 3.1.

Interestingly, the orphan rule creates a power dynamic between the upstream crate defining a concept and the downstream crate implementing it with regards to breaking changes. Consider a crate importing `StringConvertible`. This crate is allowed to define conformances for any local type—i.e., from the same crate—including generic ones. However, implementing `StringConvertible` on a foreign—i.e., from another crate—type is not allowed, even if that type is generic and instantiated with a local argument. The restriction is due to the upstream crate being allowed to add new conformances of `StringConvertible` for any type with the promise that such additions will not cause breaking changes.

■ **Listing 12**   Rust's orphan rules

```
1  struct S(String);
2  struct SBox<T>(T);
3
4  impl StringConvertible for S { ... } // <- ok
5  impl<T: StringConvertible> StringConvertible for SBox<T> { ... } // <- ok
6  impl StringConvertible for Option<S> { ... } // <- illegal definition
```

Since crates cannot form circular dependencies in Rust, additional models provided upstream cannot overlap with any existing model downstream, with one notable exception. A model defined for a universally quantified variable—i.e., a *blanket impl* in Rust parlance—can create downstream overlaps. Listing 9 presented an example in Haskell that we repeat in Rust here:

```
1  impl<T: Iterator> StringConvertible for T where <T as Iterator>::Element: StringConvertible { ... }
```

This definition will overlap with any downstream definition providing a conformance to `StringConvertible`. To fend off conflicts, Rust therefore requires that the author of the upstream crate chooses whether or not they wish to include a blanket impl before publishing their crate. Adding one later will be treated as a source-breaking change. Note that it is not possible to define more than one blanket impl, as per the rules discussed in Section 3.1.

Polymorphic traits bring other complications. Suppose we want to convert `S` to a `String` using Rust's conversion trait, `From<T>`, mentioned in Section 2.3:

```
1  impl From<S> for String { fn from(x: S) -> String { x.0 } }
```

Somewhat surprisingly, this definition satisfies the orphan rules despite being an implementation of a foreign trait `From` for a foreign type `String`. Given a foreign polymorphic trait, the orphan rules accept models for a foreign type if a local type occurs in the generic arguments of the trait. That is the case here since `From` is applied to the local type `S`. In addition, all types occurring in the model's signature have to be *covered*: they cannot be a blanket type parameter on their own. Otherwise, incoherence creeps back. Should the following situation be accepted, for example, a crate importing both *B* and *C* would obtain conflicting conformances of `From<B>` for `C`:

```
1  // In crate B
2  struct B;
3  impl<T> From<B> for T { ... } // <- illegal definition
4
5  // In crate C
6  struct C;
7  impl<T> From<T> for C { ... } // <- ok
```

We sympathize with readers who find these rules confusing. While Rust handles orphan instances much more effectively than Swift and Haskell, this section demonstrates that it is at the cost of significant complexity.

### 3.2.2 Unsoundness in Swift

As mentioned in Section 3.1, Swift checks for overlapping models at the point of definition rather than the point of use. Implicit resolution then relies on this property to assume that no ambiguity can ever occur. This strategy is sound in the context of a single compilation target. However, as Swift does not enforce restrictions analogous to Rust's orphan rule, one can violate the assumption of model uniqueness in a program containing multiple compilation targets.

We have been able to exploit this violation to expose an unsoundness in Swift's type system. The demonstration emulates the setup shown in Figure 1, with *A*1 and *A*2 merged into a single module *A*. The contents of each module is shown below.

```
1  // In module A
2  public protocol P { associatedtype X }
3  public struct S<T: P> {
4    public let x: T.X
5    public init(_ x: T.X) { self.x = x }
6  }
```

```
1  // In module D
2  import A; import B; import C
3
4  print([
5    B.mk() as! S<Int>, A.mk() as! S<Int>
6  ])
```

```
1  // In module B
2  import A
3  extension Int: P { public typealias X = Bool }
4  public func mk() -> Any { S<Int>(false) }
```

```
1  // In module C
2  import A
3  extension Int: P { public typealias X = Int }
4  public func mk() -> Any { S<Int>(42) }
```

As of version 6.0, Swift accepts this program despite the incoherence caused by the two orphan instances from modules $B$ and $C$. Due to this incoherence, the compiler cannot possibly determine the correct memory layout allocating an array of S<Int>. Using the conformance from $B$, each instance of A<Int> is expected to occupy one byte of storage, enough to store a Boolean. Using the conformance from $C$, each instance is expected to occupy one word of storage, enough to store an integer. The program is clearly unsound. Through B.mk and C.mk, $D$ creates two instances of A<Int> with different sizes, wrapped in existential containers [19], which it unwraps to form an array. The resulting value is nonsensical.

Incoherence alone is not responsible for the unsoundness; two other key components are at play. First, the conflicting conformances of Int to P specify distinct associated types. Second, the contents of A<Int> depends on the conformance that has been used to create an instance. Thus, the expression "A<Int>" does not carry enough information to describe instances built in different contexts. Unsoundness disappears if either of these components are removed from the picture. Uses of A<Int>'s methods could still behave differently depending on their receiver's origin but it would no longer be possible to confuse the type system about the type of a particular value.

The program no longer compiles if we dispense with existential containers and return S<Int> rather Any from B.mk and C.mk. Interestingly, though, the error occurs at linking time, after the program has successfully type checked. As it was later confirmed in an exchange with a former member of the compiler team, this behavior shows that Swift's intermediate representation contains sufficient information to distinguish between type applications having "captured" different conformances. Specifically, A<Int> has different mangled names in $B$ and $C$. Unfortunately, the surface syntax cannot express this difference. Neither Haskell nor Rust has surface syntax to make the distinction either. Nonetheless, the problem cannot be reproduced in those languages because Haskell's lazy overlapping check will catch the incoherence in $D$ while Rust will reject $B$ and $C$ with its orphan rules.

## 4    Coherence without Model Uniqueness

Let us now look at another approach to uphold coherence in the context of implicit programming, without resorting to global uniqueness. As mentioned in Section 2, dependently typed languages support a more flexible approach to implicit programming, thanks to their ability to express models as first-class values. Consequently, two overlapping conformances of a type to a concept can be distinguished by the values upon which they depend.

Scala exploits this mechanism to implement its support for type classes. A central feature of the language is *path-dependent typing* [1], a form of dependent typing where the definition of a type can depend on a *path*, i.e., a qualified name identifying a specific value from the context. A typical introductory example is presented below.

```
1  case class Dog(name: String):
2    class Toy
3    def toy = new this.Toy
4    def play(t: this.Toy)
5
6  val leo = Animal("Leo")
7  val rex = Animal("Rex")
8  leo.play(leo.toy) // <- ok
9  leo.play(rex.toy) // <- error; found 'rex.Toy', required 'leo.Toy'
```

The class `Dog` defines an inner class denoting the type of toy with which a dog may play. Because this type member is path-dependent, an instance of `Dog` may only play with its corresponding toy, as the error in line 9 demonstrates. Indeed, the error message reveals that the type leo.Toy is not equal to the type rex.Toy; as each of these types depend on different paths.

One can leverage this mechanism to implement concepts and their associated types in Scala. For example, the program below illustrates our running `Iterator` example:

■ **Listing 13**   The `Iterator` concept in Scala

```
1  trait Iterator[Self]:
2    type Element
3    extension (self: Self) def next: Option[(Element, Self)]
4
5  given Iterator[Long]:
6    type Element = Byte
7    extension (self: Long) def next: Option[(Byte, Long)] = self match
8      case 0 => None
9      case n => Some((n & 0xff).toByte, n >> 8)
```

The concept is expressed as a trait with a type member denoting its associated type and an extension method representing its single requirement. A model of this concept is nothing more than an instance—in the sense of object-oriented programming—of the trait. Defining it with `given` introduces it in the implicit scope so that compiler may supply it automatically to functions accepting implicit parameters:

```
1  def fold[T: Iterator, U](xs: T, x: U)(f: (U, T.Element) => U): U =
2    xs.next match
3      case Some((y, ys)) => fold(ys, f(x, y))(f)
4      case None => x
5
6  val _ = println(fold(0x2a2aL, 0)(_ + _))
```

Note that nothing in the two above examples is specific to type classes. One could also have created an instance of `Dog` and add it to the implicit scope, the same way as we did for implementing a conformance of `Long` to `Iterator`.

The unsoundness issue that we observed in Swift is avoided in Scala, thanks to path-dependent types. Consider the following snippet, which is similar to the problematic code in module *D*:

```
1  trait P[Self] { type X }
2  class S[T: P](val x: T.X)
3
4  object B:
5    given m: P[Int] with { type X = Boolean }
6    def mk = S(false)
7
8  object C:
9    given m: P[Int] with { type X = Int }
10   def mk = S(42)
11
12 object D:
13   val notInvalid = List(B.mk.x, C.mk.x)
```

Perhaps surprisingly, no error is reported. Upon closer inspection, we observe that the inferred type of `notInvalid` is `List[B.m.X | C.m.X]`, that is a list of the union of two type member selection. This type is the most precise one we could expect. Further, the path-dependency accurately accounts for the different witnesses of `Int`'s conformance to `P` that have been used to build those values. This mechanism lets us reason about different conformances within one single scope without inconsistencies and without the need for convoluted restrictions, such as Rust's orphan rules. Consequently, we are able to depend on multiple libraries with potentially various implementations of the model of a type to some type class without having two worry about importing incoherent combinations of instances. In fact, we are free to provide as many overlapping instances as we like, even within a single local scope, since the soundness of the system does not rely on the model uniqueness.

We gained flexibility by being able to express the prefix from which associated types were selected, which is possible in any language with some form of dependent typing. In contrast, the previous approach had to enforce that any selection for a given concept and associated type had to be made on some unique witness. We observe that dependently-typed and non-dependently-typed languages rely on prefix selections and global uniqueness respectively as sufficient—but perhaps not necessary—conditions to uphold coherence.

Finally, we remark that the expressiveness of dependent types comes with some loss of equational reasoning. For example, given two types `T` and `U` with conformance's to `P`, we can no longer conclude from the fact that `T` is equal to `U` that `T.X` is equal to `U.X`. Indeed, such a deduction would be incorrect in general in the presence of overlapping local instances. Our previous example illustrates this situation: the two conformances are `B.m` and `C.m`, which both witness the conformance of the same type to `P`, yet `B.m.X` is not equal to `C.m.X`.

## 5 Conclusion

We investigated four popular programming languages, Rust, Swift, Haskell and Scala, on the problem of upholding coherence of type class resolution. The former three employ global model uniqueness as a common answer to coherence. However, we showed in Section 3.1 that this approach is difficult to uphold and ultimately requires over-approximation. Furthermore, we showed in Section 3.2 that, even if perfect overlapping detection was possible, global uniqueness causes major headaches to modular software design.

We described each language's attempt to relax the conservative restrictions imposed to guarantee global uniqueness. The state of the art is rather unappealing: overlap detection leads to convoluted yet rather restrictive systems, as we demonstrated with several examples. Neither Haskell nor Swift address the problem of maintaining coherence across libraries. While Rust's orphan rules prove to be effective, they give rise to intricate and somewhat arbitrary tweaks to balance the freedom of model implementations for both trait definers and their users. Our study also revealed an unsoundness in Swift, which will certainly get fixed, but likely by introducing yet another set of rules to its type system, akin to Rust's orphan rules.

These observations suggest that global model uniqueness is not the right approach to support a programming style based on type classes. We speculate the issue stems from the fact that this strategy, though appealing on paper, does not scale to real programming languages, as evidence by Rust, Swift, and Haskell's schemes to relax it. Unfortunately, these schemes only marginally improve on expressiveness while potentially opening avenues for unsoundness. Ultimately, detecting overlapping models, be it at definition site or use site, is a hopeless battle against undecidability.

Finally, we compared global uniqueness with a more flexible approach to coherence using dependent types, as supported by Scala. Implicit resolution is merely a way to omit certain arguments of arbitrary types, dropping uniqueness restrictions. The system then trivially addresses the issues we discussed with respect to modular design and benefits the full flexibility and soundness properties of a powerful dependently-typed language. In exchange, the language involves intricate disambiguation policies and lack completeness guarantees. Furthermore, the dependent typing has complexity costs that must be carefully weighed in the design of the programming language.

While it appears that all main stream non-dependently typed languages rely on uniqueness of type class instances for coherence, we point out that it is not a fundamental requirement. Hence, we believe that there exist a design space to explore for support of coherent resolution with locally scoped and overlapping model definitions without dependent types.

## About the authors

**Dimi Racordon** is a post-doctoral researcher at EPFL, Switzerland, working on language design and type systems for safe and efficient generic programming. Contact her at dimitri.racordon@epfl.ch.
https://orcid.org/0000-0003-0299-3993

**Eugene Flesselle** is a master student at EPFL, Switzerland. Contact him at eugene.flesselle@epfl.ch.
https://orcid.org/0009-0003-7545-594X

**Cao Nguyen Pham** is a PhD student at EPFL, Switzerland. Contact him at nguyen.pham@epfl.ch.
https://orcid.org/0009-0005-2543-3309